# The high-frequency dynamics of domain walls with strong Dzyaloshinskii-Moriya interaction

Yue Zhang[1], Mao-Kang Shen[1,&], Zai-Dong Li[2,3], Xiao-Fei Yang[1], Long You[1*]

1. School of Optical and Electronic Information, Huazhong University of Science and Technology, Wuhan, 430074, PR China.

2. Department of Applied Physics, Hebei University of Technology, Tianjin 300401, PR China.

3. Key Laboratory of Electronic Materials and Devices of Tianjin, School of Electronics and Information Engineering, Hebei University of Technology, Tianjin 300401, PR China

E-mail: [lyou@hust.edu.cn](lyou@hust.edu.cn) (Long You)

&: This author has the same contribution with Yue Zhang

**Abstract**

Domain walls (DWs) in perpendicularly magnetized nanotracks (PMNTs) with interfacial Dzyaloshinskii-Moriya interaction (DMI) have become the primary objects of theoretical and experimental interest due to their technological suitability in spintronic nanodevices. Chiral DWs in PMNTs can be driven efficiently by the spin-orbit torque. However, the high-frequency dynamic behavior of the chiral DW has not been explored. In this work, using micromagnetic calculation, we have discovered a novel dynamic mode, the *sway* mode, of DWs under an out-of-plane high-frequency alternating current (AC) magnetic field in a PMNT with strong DMI. This dynamic phenomenon is strictly related with DMI-related boundary effect and can be understood in terms of the propagation of an amplitude-tuned spin wave in the DW plane. The spin wave exhibits some characteristic frequencies due to the space-confinement of DW. This work offers the possibility of a visual route for characterizing DMI.

**Introduction**

Understanding and manipulation of domain walls (DWs) motion behavior in ferromagnetic nanotracks have been extensively investigated over the last decades, driven by academic interest and the possibility of applications in high-performance low-dissipation spintronic devices. The DW motion can be controlled by static or alternating magnetic field [1–4], spin transfer torque (STT) [5–12], spin-orbit torque (SOT) [13-25], spin waves (magnons) [26-31], etc. Recently, several theoretical and experimental studies have revealed that the interfacial Dzyaloshinskii-Moriya interaction (DMI) plays a crucial role in the formation of a chiral Néel-typed DW in heavy metal (HM)/ferromagnet (FM) heterostructures [15-17, 32]. In such heterostructure, the DW can be driven efficiently by the current-induced SOT arising from spin Hall effect and/or Rashba spin-orbit coupling [14, 20, 33]. Additionally, intriguing theoretical and experimental investigation reveals that the DMI has enormous potential to influence DW dynamics of chiral spin structure in PMNTs. For instance, the DW with strong DMI tilts under the action of current-induced SOT [20-25], which is attributed to the competition between DMI and SOT. The DMI tends to pin the moments to be perpendicular to the DW plane, while the SOT rotates the DW moments. As a compromise, the DW plane tilt [23]. On the other hand, the DMI-related boundary effect may be another reason for the DW tilting. It results in the curved DW plane and the difference of DW velocity between the boundary and the inner part of the track [24, 34, 35].

In addition to the current-induced motion of the chiral DW, the dynamic behavior of chiral spin textures including skyrmions or DWs under a high-frequency magnetic field is also interesting. For example, an AC magnetic field with a frequency of several GHz results in a periodic dislocation for a skyrmion in a confined geometry, such as a breathing or gyrotropic mode [36-40]. However, what if the AC magnetic field is applied to a DW with *chirality?* This question has not been explored. In this study, we discovered numerically that strong DMI could lead to a novel high-frequency sway mode for a DW with strong DMI in PMNTs under a high-frequency out-of-plane AC magnetic field. This sway characteristic is peculiar to a chiral DW with strong DMI and is attributed to an amplitude-tuned spin wave transporting in the DW plane. This finding offers the possibility of a visual route for characterizing the propagation of spin wave in a system with strong DMI.

**Results**

**1. Micromagnetic simulation investigation of the sway mode**

**1.1 The sway mode under an out-of-plane AC magnetic field**

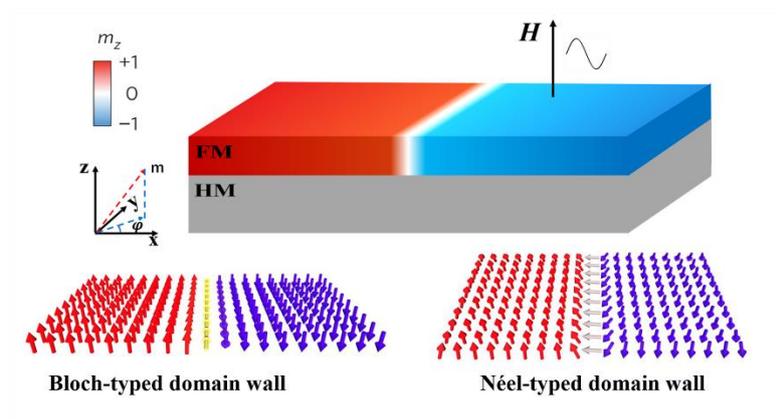

**Fig. 1. Schematic (up) of a chiral DW under an out-of-plane AC magnetic field in a FM/HM bilayer. The Bloch-typed DW is generated without DMI (down left) and the Néel-typed DW is found with none-zero DMI (down right).**

We investigate the dynamic behavior of a chiral DW in an HM/FM bilayer track under an out-of-plane AC magnetic field using the micromagnetic simulation. The FM layer with the saturation magnetization $M_S$ exhibits perpendicular magnetic anisotropy (PMA). The interfacial DMI is assumed to be generated at the HM/FM interface. When the DMI constant ($D$) is zero, the initial DW exhibits a Bloch-typed structure (down left). However, when non-zero $D$ is considered (between −0.05 mJ/m$^2$ and −1.2 mJ/m$^2$), the initial DW changes to a Néel-typed one (down right) even though the $D$ is as small as −0.05 mJ/m$^2$. Due to the negative $D$, the DW shows a left-handed chirality.

To determine the characteristic frequencies for the dynamical modes, the initial equilibrium state of DW was perturbed by an out-of-plane AC magnetic field as a sinc function of time with the frequency varying between 0 and 10 GHz. The equilibrium magnetization was removed from the recorded magnetization dynamics. The remaining time-dependent magnetization was transformed into the frequency domain using a Fast Fourier Transformation (FFT) by the spatially averaged method [36].

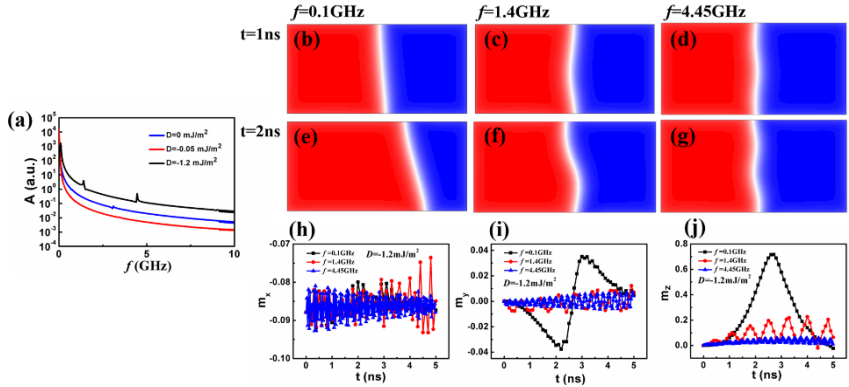

**Fig. 2. The dynamics of DW under out-of-plane AC magnetic fields. (a). The FFT spectra of the time-dependent magnetization in the FM/HM bilayer with different DMI constants. (b)~(g). The snapshots of the DWs for $D$ = −1.2 mJ/m$^2$ at 1 ns and 2 ns under a sine AC magnetic field with the characteristic frequencies in (a). (h)-(j). The time-dependent $m_x$, $m_y$, and $m_z$ for $D$ = −1.2 mJ/m$^2$ under the sine AC magnetic field with the characteristic frequencies in (a).**

A very weak peak was detected around 3 GHz and 3.4 GHz in the FFT spectra for the zero or

−0.05-mJ/m² $D$, respectively (Fig. 2(a)). However, when the $D$ is −1.2 mJ/m², three characteristic frequencies were found at 0.1 GHz, 1.4 GHz, and 4.45 GHz. The DW motion under the sine AC magnetic fields at these frequencies were analyzed individually. An oscillating forward of DW (the oscillation of $m_z$ (Fig. 2(j))) was observed, and the amplitude of the oscillation decreases with the increase in the frequency. On the other hand, $m_y$ also exhibits similar oscillation behavior (Fig. 2(i)), which indicates the precession of the moments in the DW. On the other hand, the shape of DW also depends on the frequency. When the frequency is as low as 0.1 GHz, the DW plane tilts but keeps straight, and the tilting angle oscillates (Figs. 2(b) and (e)). At the higher frequencies, the DW deforms and exhibits an S-shaped plane (Figs 2(c), (d), (f), and (g)).

**1.2 The sway mode of a moving DW under both SOT and an out-of-plane AC magnetic field**

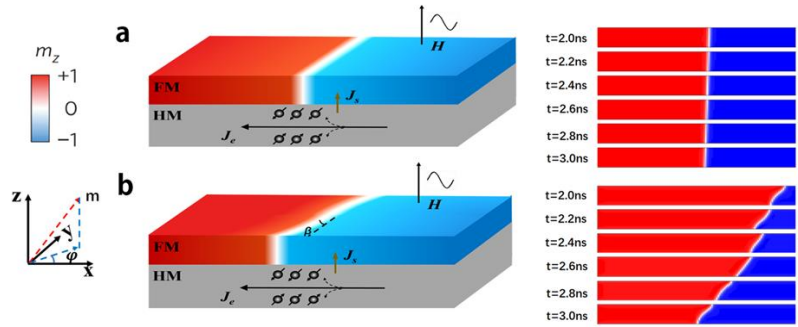

**Fig. 3. Schematic (left) and the snapshots (right) of a chiral DW motion driven by both SOT and an out-of-plane AC magnetic field in an FM/HM bilayer with (a) a very weak negative interfacial DMI ($D$ = −0.05 mJ/m²), and (b) a strong DMI ($D$ = −1.2 mJ/m²). The current density in both cases is 5×10¹¹ A/m². ($t$ is from 2 ns to 3 ns, a period of the AC magnetic field when the motion of DW becomes stable.)**

We also investigated the dynamics of DW under both SOT and the AC magnetic field. Since the Bloch DW without DMI cannot be driven by SOT, we focused on the motion of DW with DMI. On the right of Fig. 3 there are representative snapshots of the DW structures in a period of the AC field with the magnitude ($H_m$) of 50 mT and the frequency ($f$) of 1 GHz. When the current density ($J$) is $5 \times 10^{11}$ A/m$^2$ and a 1-GHz out-of-plane AC field ($H_m$ = 50 mT) is applied to the DW with $D$ = −0.05 mJ/m$^2$ (Fig. 3(a)), the DW moves to the left slowly. In this case, neither tilting nor sway was observed. However, when the $D$ is −1.2 mJ/m$^2$, the DW moves to the left with a clear tilting at a much higher speed, and the sway of the DW plane was observed. In the oscillation of the AC magnetic field, the DW plane is curved and swaying, and the deformation occurs periodically.

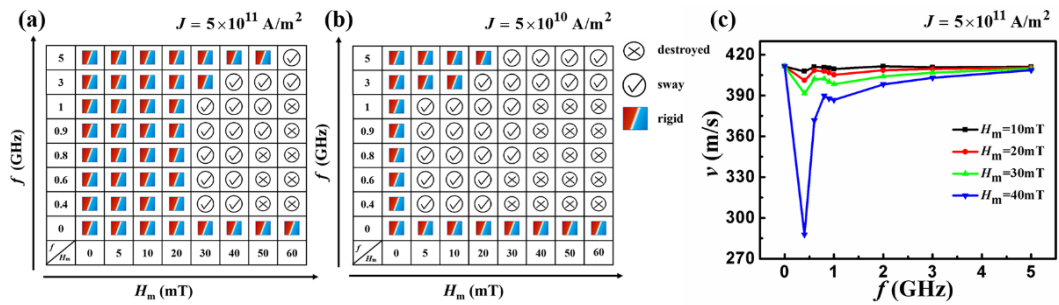

**Fig. 4.** The phase diagrams for the morphologies and the velocity of the DW: The phase diagrams for the DW with $D$ = −1.2 mJ/m$^2$ under the AC field with $f$ varying from 0 to 5 GHz and $H_m$ changing from 0 to 100 mT when (a) $J = 5 \times 10^{11}$ A/m$^2$, and (b) $J = 5 \times 10^{10}$ A/m$^2$. (c). The velocity of the DW as a function of $f$ under an ac magnetic field with the amplitudes varying between 10 mT and 40 mT ($J = 5 \times 10^{11}$ A/m$^2$).

The sway of the DW with a strong DMI ($D$ = −1.2 mJ/m$^2$) is significantly influenced by tuning the current density and the AC field. According to the phase diagram in Fig. 4(a) ($J = 5\times10^{11}$ A/m$^2$), when $H_m$ is 20 mT or weaker, the DW is rigid in the whole $f$ range (0—5 GHz, the 0 GHz means without external field). When $H_m$ increases to 30 or 40 mT, the sway of DW was observed when the $f$ is several hundred MHz, which is in the range of the frequency for the DW resonance. Under a stronger AC field, the $f$ for the sway shifts to a higher value, and the DW is destroyed at higher $f$. As shown in Fig. 4(b) ($J = 5\times10^{10}$ A/m$^2$), the reducing of $J$ leads to the decrease of $f$ and

$H_m$ for the sway mode and the destroy of DW. The sway mode is found when the $f$ and $H_m$ are as low as 0.4 GHz and 5 mT, respectively. This result indicates that the AC field for exciting the sway mode competes with SOT. SOT needs to be reduced so that the sway can occur under a low-frequency weak AC field.

Because of the background from the unidirectional motion of DW, it is difficult to determine the characteristic frequency by FFT. However, the characteristic frequency can also be reflected from the frequency-dependent velocity of DW driven by both SOT and AC field (Fig. 4 (c)). Significant reducing of DW velocity occurs in the frequency range of hundreds of MHz, which is the frequency range for the observed sway mode of the moving DW. Much weaker velocity peaks appear around 1 GHz where the sway mode is observable only if the amplitude of AC field is strong enough (Fig. 4(a)). When the frequency is higher than 1 GHz, the velocity of DW is recovered to that without any external magnetic field.

## 2. Investigation about the sway mode of DW using the collective coordinate method (CCM)

### 2.1. The CCM for the sway mode under an out-of-plane AC magnetic field

By assuming the DW plane is *straight*, the dynamics of DW can be depicted by a group of Thiele equations for several collective coordinates, such as the central position $q$, the azimuthal angle $\varphi$ for the moment in the central of DW, and the tilting angle $\beta$ of the DW plane, etc.

When only the out-of-plane AC magnetic field is applied, the Thiele equations for $q$ and $\varphi$ are derived (the method for the derivation is shown in the Appendix.) as:

$$\frac{\alpha}{\Delta}\dot{q} + \dot{\varphi} = \gamma_0 H_z \tag{1},$$

$$\frac{\dot{q}}{\Delta} - \alpha\dot{\varphi} = -\frac{\gamma_0 \pi D \sin\varphi}{2\Delta M_S} - \frac{\mu_0 \gamma_0 N_x M_S \sin(2\varphi)}{2} \tag{2}.$$

Here α, γ₀, μ₀, $N_x$, Δ, and $H_z$ are the damping coefficient, the gyromagnetic ratio of an electron, the permeability of vacuum, the demagnetization factor, the DW width, and the external AC field along the z direction. The $H_z$ is expressed as

$$H_z = H_m \sin(2\pi f t) \quad (3).$$

When an out-of-plane field is applied, the Δ does not change and it can be taken as a constant.

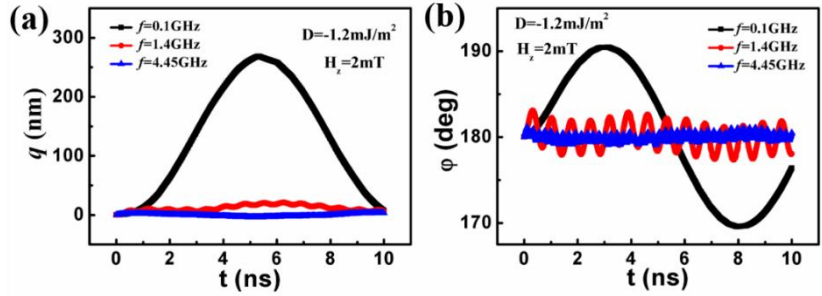

**Fig. 5.** (a) The central position *q* and (b) the azimuthal angle *φ* of DW calculated by the Eqs. (1) and (2) under the 2-mT out-of-plane AC fields with different frequencies ($D = -1.2$ mJ/m²).

The group of equations (Eqs. (1) and (2)) was solved numerically using the 4th-order Runge-Kutta algorithm. As shown in Fig. 5, at the frequencies of 0.1, 1.4, and 4.45 GHz, the characteristic frequencies in the FFT spectra (Fig. 2), the changes of *q* and *φ* with time contributes to the temporal $m_z$ and $m_y$ in Fig. 2, respectively.

**2.2. The CCM for the sway mode under both SOT and an out-of-plane AC magnetic field**

The DW motion driven by both external magnetic field and SOT was also described by CCM. When the DMI is strong, in addition to *q* and *φ*, the tilting angle *β* of DW (depicted in Fig. 3) also needs to be considered [23]. The group of Thiele equations for *q*, *φ*, and *β* were theoretically derived

$$\frac{\alpha\cos\beta}{\Delta}\dot{q}+\dot{\varphi}=\frac{\pi\gamma_0 H_{SO}J\cos\varphi}{2}+\gamma_0 H_z \quad (4),$$

$$\frac{\cos\beta}{\Delta}\dot{q}-\alpha\dot{\varphi}=\frac{\gamma_0\pi D\sin(\varphi-\beta)}{2\Delta\mu_0 M_s}+\frac{\gamma_0 N_x M_s \sin[2(\varphi-\beta)]}{2} \quad (5),$$

$$\frac{\pi^2\Delta\alpha\mu_0 M_s}{6\gamma_0}[\tan^2\beta+(\frac{w}{\pi\Delta})^2\frac{1}{\cos^2\beta}]\dot{\beta}=-\{\Delta\mu_0 N_x M_s^2 \sin[2(\varphi-\beta)]$$
$$+\Delta\mu_0 N_x M_s^2 \cos^2(\varphi-\beta)\tan\beta+\frac{4A}{\Delta}\tan\beta-\frac{\pi D\sin\varphi}{\cos\beta}\} \quad (6).$$

Here $w$ and $H_{SO}$ are the width of track and the effective field for SOT, respectively.

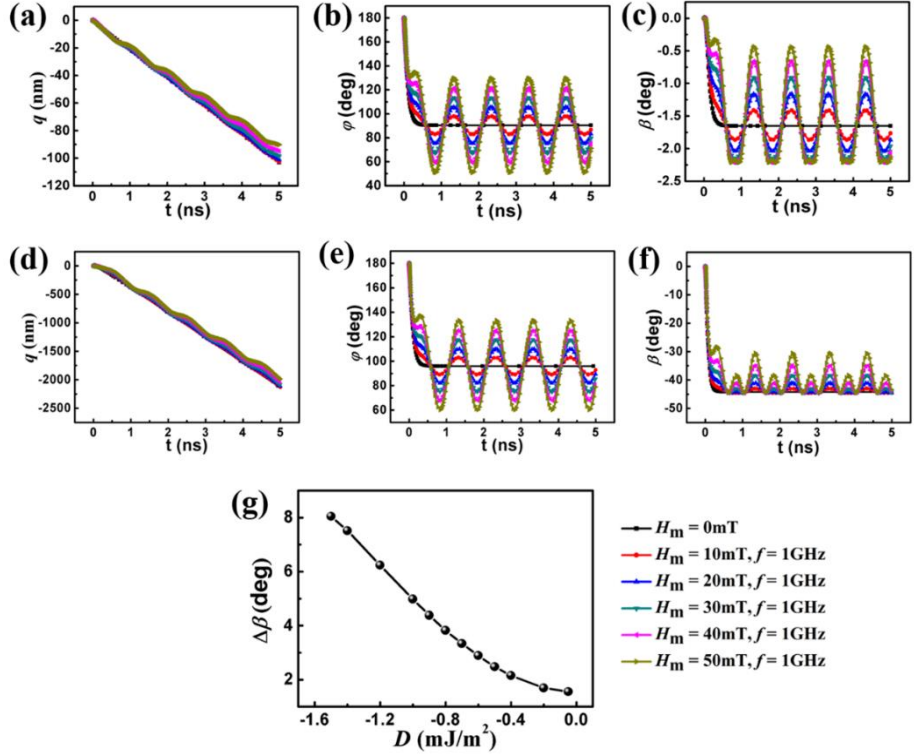

**Fig. 6.** (a), (d) The central position $q$, (b), (e) the azimuthal angle $\varphi$, and (c), (f) the tilting angle ($\beta$) of DW calculated by the Eqs. (4)—(6) under the 1-GHz AC fields with different $H_m$. The $D$ are −0.05 mJ/m² ((a)-(c)) and −1.2 mJ/m² ((d)-(f)), and the current density $J$ is fixed at 5×10¹¹ A/m². (g). The difference of the tilting angles in the two semi-cycles of $\varphi(t)$ as a function of $D$.

The group of equations (Eqs. (4)—(6)) was also solved numerically using the 4$^{th}$-order

Runge-Kutta algorithm. As shown in Fig. 6(a), for both $D$ (−0.05 mJ/m$^2$ and −1.2 mJ/m$^2$), the application of the AC magnetic field along the $z$ direction and SOT results in the oscillating forward of DW with a small reducing of velocity when compared to that without any external field. The velocity of DW with the weaker DMI is significantly smaller than that for the larger one. The application of the AC field also leads to the oscillation of $\varphi$ around the equilibrium $\varphi$ under the zero field, and the amplitude of the oscillation increases with the enhancement of $H_m$ (Fig. 6(b) and (e)). The difference in $D$ makes little impact on the oscillation of $\varphi$ except that the equilibrium $\varphi$ for the smaller $D$ is closer to 90° due to the smaller DMI energy barrier for the rotation of moment under SOT.

The tilting angle $\beta$ also oscillates. However, the $\beta$ for the smaller $D$ is only around 1.5°, which is significantly smaller than that of the larger $D$ (around 45°). In addition, there are two distinct amplitudes of $\beta(t)$ corresponding to the two semi-cycles of $\varphi(t)$. The larger amplitude is related to the positive semi-cycle of $\varphi(t)$, while the smaller one corresponds to the negative semi-cycle of $\varphi(t)$. When the $D$ is as small as −0.05 mJ/m$^2$, the two different $\beta(t)$ distribute on the two sides of the equilibrium value. When the $D$ is as large as −1.2 mJ/m$^2$, both $\beta(t)$ appear on the same side of the equilibrium $\beta$. The difference of $\beta$ for the two semi-cycles is defined as:

$$\Delta\beta = \begin{cases} \max(\beta(0,T/2)) - \max(\beta(T/2,T)) & D < D_c \\ \max(\beta(0,T/2)) - \min(\beta(T/2,T)) & D \geq D_c \end{cases} \quad (7).$$

Here $T$ is the period, and $D_c$ is the critical $D$ for reversing the direction for the tilting during $T/2$ and $T$. A monotonous increase of the $\Delta\beta$ with $D$ is shown in Fig. 6(g). This offers the possibility of a visual route for measuring DMI.

The CCM analysis indicates that the application of a high-frequency AC magnetic field in the $z$ direction leads to the precession of moments in the DW and the oscillation of the tilting angle of the DW with a strong DMI. From the physics point of view, the DMI is equal to an effective magnetic field in the DW region that tends to pin the moments to be perpendicular to the DW plane [23]. Therefore, under an AC field, the whole DW may sway following the precession of the DW moments. In addition, because of the chirality of the DW as a result of the unidirectional DMI effective field, the moments at the two sides of the equilibrium orientation of $\varphi$ (the positive and the negative semi-cycle of $\varphi(t)$) have different energies. This energy difference enlarges with the increase in $D$. As a result, in a period of $\varphi(t)$, the amplitudes of $\beta$ have distinct values.

## 3. The physical mechanism for the sway mode

The above CCM analysis is based on the assumption that the DW plane is straight. It is not able to reveal the reason for the deformation of the DW plane quantitatively. In the following, more details about this DW dislocation are discussed.

### 3.1 The propagation of an amplitude-tuned spin wave along the DW plane with strong DMI

The moments in the DW can be driven to precess under the combined action of the DMI effective field along the $x$ directions and the out-of-plane AC magnetic field. Since the DMI induces different orientation of moments at the boundary and in the inner part of the track [41], the moment precession along the DW plane is not uniform. This can be clearly seen from Fig. 7. During a period of the AC field, the moments in the DW experience precession for both $D$. However, for the small $D$ (−0.05 mJ/m$^2$), the shape of the DW plane keeps straight (Fig. 7(a)). The amplitude of $\varphi$ is smaller than 10, and that near the track edge is a little smaller than that in the inner part by around 2 degrees (Fig. 7(c)). Nevertheless, when $D$ is −1.2 mJ/m$^2$, the amplitude of $\varphi$ is as large as about 40 degrees, and that near the track boundary is larger than that in the inner part by several degrees (Fig. 7(d)). This indicates that the moment precession near the track boundary becomes dominant over that in the inner part of the track when the DMI is strong enough, which is originated from the DMI-related boundary effect [41].

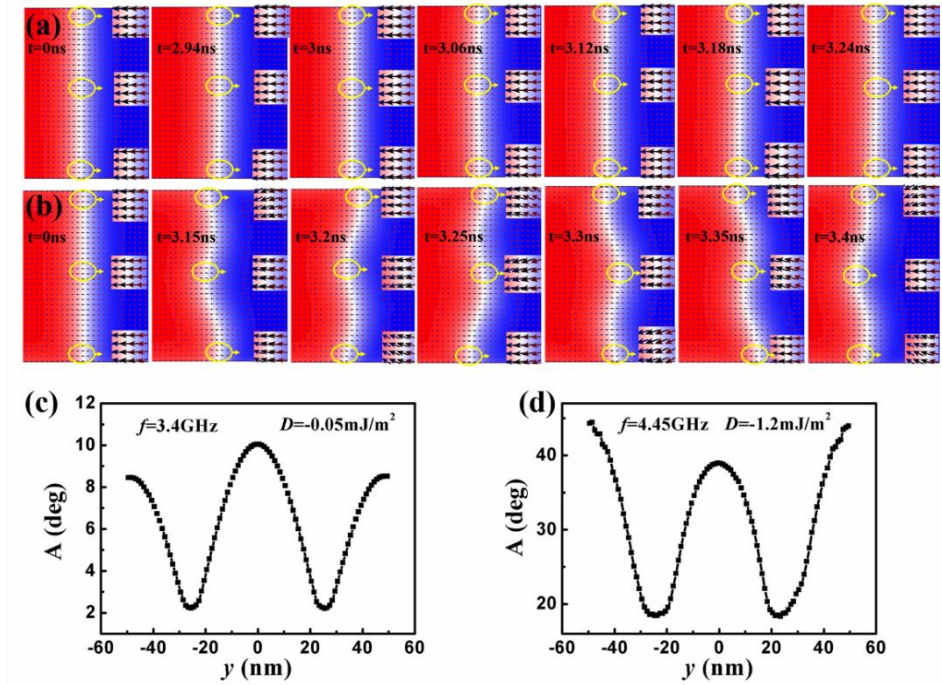

Fig. 7. The moment distribution in the DW, driven by both AC magnetic field for (a) $D = -0.05$ mJ/m$^2$ and (b) $D = -1.2$ mJ/m$^2$ at the initial state ($t = 0$ ns) and during a period of the AC field. The variation of amplitude of $\varphi(t)$ along the track width direction (the y direction) for (c) $D = -0.05$ mJ/m$^2$ and (d) $D = -1.2$ mJ/m$^2$.

The non-uniform precession of moments excites a spin wave transporting along the DW plane. The spin wave can be characterized from the $\varphi$ as a function of time and space (Fig. 8). When the $D$ is $-0.05$ mJ/m$^2$, the moments at both track boundaries and in the inner part of the DW plane precess as a sine function of time. The phases of $\varphi(t)$ at the two track boundaries are the same but opposite to that in the central of the DW (Fig. 8(a)). This means the transport of two spin waves towards opposite directions in the upper and lower parts of the DW plane. However, the amplitude of $\varphi(t)$ throughout the track region is smaller than 10 degrees, indicating the spin wave is quite weak. When the $D$ is −1.2 mJ/m², however, a monotonous phase shift is found from one boundary of the track to the other (Fig. 8(b) and (c)). This indicates the unidirectional transport of spin wave along the DW plane. When the frequency is 1.4 GHz, a beat frequency is also found, showing the combination of several precessions with close frequencies. Yet, when the frequency is as high as 4.45 GHz, the precession becomes a harmonic oscillation with stable amplitude and frequency.

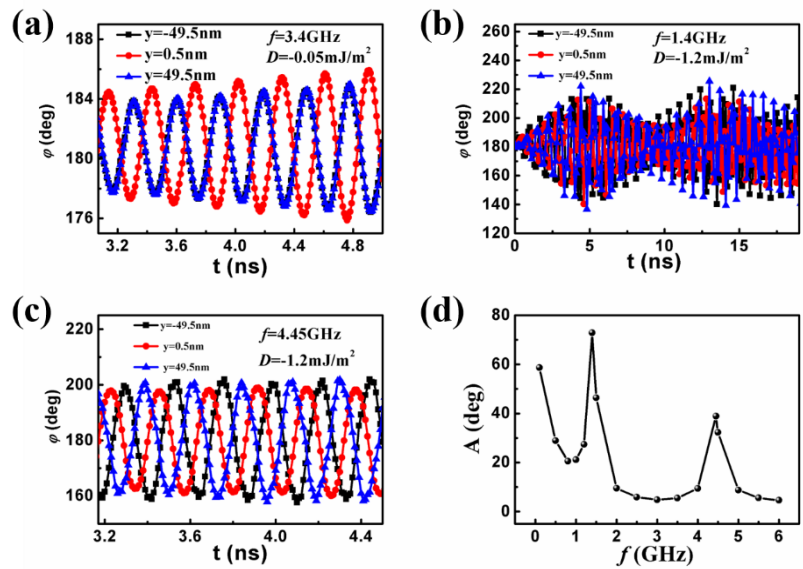

**Fig. 8.** The time-dependent $\varphi$ for (a) $D$ = −0.05 mJ/m² and $f$ = 3.4 GHz, (b) $D$ = −1.2 mJ/m² and $f$ = 1.4 GHz, (c) $D$ = −1.2 mJ/m² and $f$ = 4.45 GHz, and (d) the frequency dependence of the amplitude for $D$ = −1.2 mJ/m².

Fig. 8(d) shows the frequency-dependence of the amplitude of precession in the central of the DW. Three peaks of amplitude around 0.1 GHz, 1.4 GHz, and 4.45 GHz were observed, which is consistent well with the FFT results in Fig. 2. In general, a series of discrete frequencies is the typical properties for a wave propagating in a space-confined region, such as the wave guide.

When the frequency of external magnetic field is equal to the characteristic frequency, the precession is significantly enhanced.

**3.2 The revised CCM by considering the space distribution of the collective coordinates.**

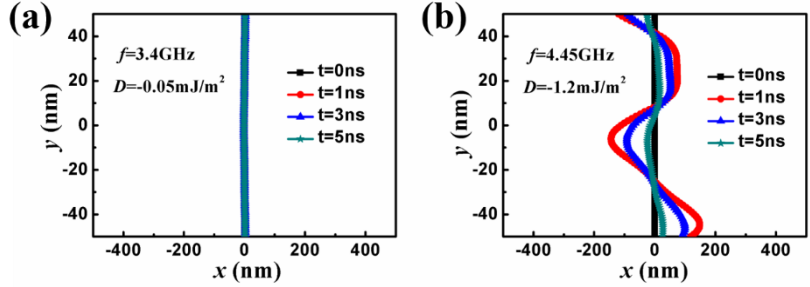

**Fig. 9. The time-dependence of DW shape for (a) $D = -0.05$ mJ/m² and $f = 3.4$ GHz and (b) $D = -1.2$ mJ/m² and $f = 4.45$ GHz based on the revised CCM by considering the space distribution of moments in the DW plane.**

The results in 3.1 reveal the transport of an amplitude-tuned spin wave in the DW plane with DMI under an out-of-plane AC magnetic field. Since the previous CCM assumes the DW plane is straight, the deformation of DW cannot be reflected. In this part, we have revised it by considering the space distribution of collective coordinates for showing the deformation of DW.

The angle $\varphi(t)$ in the previous CCM is expanded to $\varphi(y, t)$ to consider the variation of $\varphi$ across the track width. Based on the spin wave revealed in the 3.1, the $\varphi(y, t)$ can be expressed as:

$$\varphi(y,t) = A(y)\sin(2\pi ft + \phi(y)) \tag{8}$$

For $D = -0.05$ mJ/m², the $A(y)$ and $\phi(y)$ is assumed to be

$$A(y) = \frac{\pi}{180}(6.5 + 3.5\cos(\frac{\pi y}{25})) \tag{9},$$

and

$$\phi(y) = \left|\frac{\pi y}{50}\right| \tag{10}.$$

When $D = -1.2$ mJ/m², the $A(y)$ and $\phi(y)$ is assumed to be

$$A(y) = \frac{\pi}{6} + \frac{\pi}{15}\cos(\frac{\pi y}{25}) \tag{11},$$

and

$$\phi(y) = \frac{-3\pi y}{100} \tag{12}.$$

Using Eqs. (8)~(12) and the principle of CCM, the $q(t)$ can also be expanded to $q(y, t)$ that satisfies

$$\dot{q}(y,t) = \frac{\Delta}{1+\alpha^2}(-\frac{\gamma_0 \pi D \sin\varphi}{2\Delta M_s} - \frac{\gamma_0 \mu_0 N_x M_s \sin\varphi\cos\varphi}{\alpha} + \alpha\gamma_0 H_z) \tag{13}.$$

Based on the numerical solution of Eq. (13), the curve for the central line of the DW plane can be drawn (Fig. 9). When the $D$ is very small, the DW plane keeps straight, but for $D = -1.2$ mJ/m², significant deformation of the DW was observed.

The difference between the result in Fig. 9(b) and that by simulation is attributed to the simplified assumption without considering the $x$-dependence of $\varphi$ and the simple sine function for the variation of the amplitude with $y$. However, the physical figure of the sway mode has been clear enough. When the DMI is strong, the high-frequency out-of-plane AC magnetic field excites the unidirectional propagation of spin wave along the DW plane with some characteristic frequencies. The amplitude of this spin wave is large and tuned in its propagation process because of the DMI-related boundary effect, resulting in the sway of DW.

**Conclusion**

The sway mode has been revealed numerically in the chiral DWs with strong DMI under an out-of-plane AC magnetic field. The sway becomes outstanding at several characteristic frequencies and can be manipulated by changing the amplitude of external AC field. The sway is originated from the transport of an amplitude-tuned spin wave along the DW plane excited by the AC field. When the DMI is strong, the amplitude of the precession near the track boundaries is obviously larger than that in the inner part and the phase for the precession also shifts monotonously from one boundary to the other one. This can be related to the DMI-relevant boundary effect. This work offers an intuitive route for characterizing DMI.

**Appendix**

**Collective coordinate method (CCM)**

A two-dimensional (2D) CCM was applied to quantitatively describe the dynamics of DW motion. When the DMI is strong enough, the DW is described by three collective coordinates: its central position $q$, the azimuthal angle $\varphi$ of the DW magnetization in spherical coordinates, and the tilting angle ($\beta$) of the whole DW [21, 23]. When the DMI is very weak, the $\beta$ can be neglected.

The unit vector for the orientation of magnetization is described as $\vec{m} = (\sin\theta\cos\varphi, \sin\theta\sin\varphi, \cos\theta)$. The ansatz for the DW magnetization is expressed as [21, 23]:

$$\theta = 2\arctan\{\exp[(x\cos\beta + y\sin\beta - q\cos\beta)/\Delta]\},$$

and $\varphi = \varphi(t)$, (A1)

where $t$ is time, and

$$\Delta = \sqrt{A/(K - \mu_0 M_S^2/2)},$$ (A2)

is the DW width. $M_S$, $\mu_0$, $A$, $K$, and $\theta$ are the saturation magnetization, vacuum permeability,

exchange stiffness constant, magnetic anisotropy constant, and the polar angle, respectively.

The Thiele equation for the DW motion is derived using the Lagrangian approach. Let $l$ be the Lagrangian density function that is expressed as [23],

$$l = E + (M_S/\gamma)\varphi\dot{\theta}\sin\theta. \tag{A3}$$

Here $\gamma$ is the gyromagnetic ratio of an electron, $\dot{\theta}$ is the time derivative of $\theta$, and $E$ represents the total free energy density that is composed of the exchange energy density ($E_e$), magnetic anisotropy energy density ($E_a$), demagnetization energy density ($E_d$), Zeeman energy under external magnetic field ($E_H$), and free energy density from DMI ($E_{DM}$).

These free energy densities are written as

$$E_e = A\sum_{i=x,y,z}|\nabla m_i|^2, \tag{A4}$$

where $\nabla$ is the Nabla symbol.

$$E_a = (K - \frac{1}{2}\mu_0 M_S^2)\sin^2\theta, \tag{A5}$$

$$E_Z = -\mu_0 M_S H\cos\alpha, \tag{A6}$$

where $H$ is the strength of external magnetic field, and $\alpha$ is the angle between external field and magnetiation.

$$E_d = \frac{1}{2}\mu_0 N_x M_S^2 \sin^2\theta\cos^2\varphi, \tag{A7}$$

$$E_{DM} = D[m_z(\partial m_x/\partial x) - m_x(\partial m_z/\partial x)]. \tag{A8}$$

The demagnetization factor $N_x$ is approximately estimated as [21]

$$N_x = L_z \ln 2/\pi\Delta, \tag{A9}$$

where $L_z$ is the thickness for the ferromagnetic layer.

In addition to $l$, a dissipation density function $f_d$ also needs to be included for a non-conservative system [21, 23]:

$$f_d = (\alpha M_S / 2\gamma)[d\vec{m}/dt - (\gamma_0/\alpha)H_{SO}(\vec{m}\times\vec{e}_y)]^2, \tag{A10}$$

where $\alpha$ is the damping coefficient. $H_{SO}$, the effective magnetic field for SOT, is written as

$$H_{SO} = \mu_B \theta_{SH} J / \gamma_0 e M_S L_z, \tag{A11}$$

where $\mu_B$, $J$, $\theta_{SH}$, and e are the Bohr magneton, the current density, the spin Hall angle of the HM layer, and the electron charge, respectively. The parameter $\gamma_0$ is related to $\gamma$ as $\gamma_0 = \mu_0|\gamma|$.

The Lagrangian ($L$) and Rayleigh dissipation function ($F$) were deduced by spatially integrating $l$ and $f_d$ with respect to the entire track space. The set of Thiele equations are finally determined by the Lagrange–Rayleigh equation.

**Micromagnetic simulation**

The motion of the chiral DW driven by SOT and AC magnetic field was also studied by micro-magnetic simulation via the software named "object-oriented micromagnetic framework" (OOMMF) with the code of DMI [41]. The principle of the simulation is based on solving the Gilbert equation numerically

$$\frac{\partial \vec{m}}{\partial t} = -\gamma_0 \vec{m}\times\vec{H}_{eff} + \alpha(\vec{m}\times\frac{\partial \vec{m}}{\partial t}) + \gamma_0 H_{SO}(\vec{m}\times(\vec{\sigma}\times\vec{m})), \tag{A12}$$

where $\vec{H}_{eff}$ is the effective magnetic field that is related to the total free energy density ($E$):

$$\vec{H}_{eff} = -\frac{1}{\mu_0 M_S}(\frac{\delta E}{\delta \vec{m}}). \tag{A13}$$

The parameters for the simulation are the same with that for the calculation using CCM. The dimension of the track is 2000 nm × 100 nm × 0.6 nm, and the cell size is 2 nm × 1 nm × 0.6 nm. The parameters for the calculation are $M_S = 7 \times 10^5$ A/m, $A = 1 \times 10^{-11}$ J/m, $K = 4.8 \times 10^5$ J/m$^3$, and $D = -0.05$ and $-1.2$ mJ/m$^2$. The $H_m$ varies between 0 and 100 mT, and the $f$ is between 0 and 10 GHz.

**Acknowledgements**

The authors would like to acknowledge the final support from the National Natural Science Foundation of China (Nos. 11574096, 61674062, and 61774001) and Huazhong University of Science and Technology (No. 2017KFYXJJ037).